\documentclass[12pt]{iopart}

\usepackage{txfonts}
\usepackage{graphicx}
\usepackage{subfigure}
\usepackage{dcolumn}
\usepackage{bm}
\usepackage{sidecap}
\usepackage{iopams}
\usepackage{CJK}

\begin{document}

\title {Continuous-variable QKD over 50km commercial fiber}

\author{Yichen Zhang$^{1,2}$, Zhengyu Li$^1$, Ziyang Chen$^1$, Christian Weedbrook$^3$, Yijia Zhao$^2$, Xiangyu Wang$^2$, Yundi Huang$^2$, Chunchao Xu$^2$, Xiaoxiong Zhang$^2$, Zhenya Wang$^2$, Mei Li$^2$, Xueying Zhang$^2$, Ziyong Zheng$^2$, Binjie Chu$^2$, Xinyu Gao$^2$, Nan Meng$^2$, Weiwen Cai$^4$, Zheng Wang$^5$, Gan Wang$^1$, Song Yu$^2$, Hong Guo$^{1,2}$}

\address{$^1$State Key Laboratory of Advanced Optical Communication Systems and Networks, School of Electronics Engineering and Computer Science, Center for Quantum Information Technology, Center for Computational Science and Engineering, Peking University, Beijing 100871, China}
\address{$^2$State Key Laboratory of Information Photonics and Optical Communications, Beijing University of Posts and Telecommunications, Beijing 100876, China}
\address{$^3$Xanadu, 372 Richmond St W, Toronto, M5V 2L7, Canada}
\address{$^4$China Mobile Group Guangdong Co., Ltd., Guangzhou 510623, China}
\address{$^5$Xi'an Uotocom Network Technology Co., Ltd., Xi'an 710075, China}

\ead{yusong@bupt.edu.cn, hongguo@pku.edu.cn}

\vspace{10pt}
\begin{indented}
\item[]May 2019
\end{indented}

\begin{abstract}
The continuous-variable version of quantum key distribution (QKD) offers the advantages (over discrete-variable systems) of higher secret key rates in metropolitan areas, as well as the use of standard telecom components that can operate at room temperature. An important step in the real-world adoption of continuous-variable QKD is the deployment of field tests over commercial fibers. Here we report two different field tests of a continuous-variable QKD system through commercial fiber networks in Xi'an and Guangzhou over distances of $30.02$~{\rm km} ($12.48$~{\rm dB}) and $49.85$~{\rm km} ($11.62$~{\rm dB}), respectively. We achieve secure key rates two orders-of-magnitude higher than previous field test demonstrations by employing an efficient calibration model with one-time evaluation. This accomplishment is also realized by developing a fully automatic control system which stabilizes system noise, and by applying a rate-adaptive reconciliation method which maintains high reconciliation efficiency with high success probability in fluctuated environments. Our results pave the way to deploy continuous-variable QKD in metropolitan settings.
\end{abstract}

%
%
%
%
%

\section*{Introduction}
Quantum key distribution (QKD)~\cite{Gisin_RevModPhys_2002,Scarani_RevModPhys_2009} is one of the most practical applications in the field of quantum information. Its primary goal is to establish secure keys between two legitimate users, typically named Alice and Bob. Continuous-variable (CV) QKD~\cite{Weedbrook_RevModPhys_2012,Diamanti_Entropy_2015} has attracted much attention in the past few years, mainly as it uses standard telecom components that operate at room temperature and it has higher secret key rates (bits per channel use) over metropolitan areas~\cite{Jouguet_NatPhoton_2013,Pirandola_NatPhoton_2015,Weedbrook_PhysRevA_2014,Soh_PhysRevX_2015,Qi_PhysRevX_2015,Kumar_NewJPhys_2015}. We deployed the CV-QKD protocol based on coherent states~\cite{Grosshans_PhysRevLett_2002,Grosshans_Nature_2003,Weedbrook_PhysRevLett_2004} with Gaussian modulation that has been proven secure against arbitrary attacks~\cite{Grosshans_PhysRevLett_2005,Navascues_PhysRevLett_2005, Pirandola_PRL_2008, Pirandola_PhysRevLett_2009}. Such attack is the most optimal in the asymptotical limit~\cite{Renner_PhysRevLett_2009} and is also used in the finite-size regime~\cite{Leverrier_PhysRevLett_2013,Leverrier_PhysRevLett_2015,Leverrier_PhysRevLett_2017}. Furthermore, many experimental demonstrations of CV-QKD protocol have also been achieved~\cite{Lance_PhysRevLett_2005,Lodewyck_PhysRevA_2007,Qi_PhysRevA_2007,Khan_PhysRevA_2013,Pirandola_NatPhoton_2015}.

Generally speaking, there are three basic criteria for a practical QKD system: automatic operation, stabilization under a real-world environment, and a moderate secure key rate~\cite{Diamanti_npj_2016}. Up to now, all previous long distance CV-QKD demonstrations have been undertaken in the laboratory without perturbation of the field environment~\cite{Jouguet_NatPhoton_2013,Huang_SciRep_2015}. More recently, demonstrations based on continuous-variable quantum teleportation and continuous-variable Einstein-Podolsky-Rosen entangled states under labortory environment have been reported~\cite{Wang_PRApplied_2018, Huo_SciAdv_2018}.
 The longest field tests of a CV-QKD system have been achieved over a $17.52$~{\rm km} deployed fiber ($10.25$~{\rm dB} loss)~\cite{Huang_OptLett_2016} and a $17.7$~{\rm km} deployed fiber ($5.6$~{\rm dB} loss)~\cite{Jouguet_OptExpr_2012}, where the secure key rates were $0.2$~{\rm kbps} and $0.3$~{\rm kbps}, respectively. Compared with field tests for discrete-variable QKD systems~\cite{Chen_OptExpr_2010,Sasaki_OptExpr_2011,Stucki_NewJPhys_2011,Wang_OptExpr_2014,Shimizu_JLightwaveTechnol_2014}, these demonstrations have limited transmission distances and low key rates. The demonstration of field tests over longer metropolitan distances using CV-QKD has yet to be achieved.

There are several challenges to develop a practical CV-QKD system from a laboratory to the real world. Deployed commercial dark fibers are inevitably subject to much stronger perturbations due to changing environmental conditions and physical stress. This in turn causes disturbances of the transmitted quantum states. Deployed commercial dark fibers also suffer from higher losses due to splices, sharp bends and inter-fiber coupling. The software and hardware of CV-QKD modules must not only be designed to cope with all the conditions affecting the transmission fiber, but also must be robustly engineered to operate in premises designed for standard telecom equipment. Furthermore, as the system need to run continuously and without frequent attention, they should also be designed to automatically recover from any errors and shield the end users from service interruptions.

In this paper, we deploy CV-QKD protocol based on coherent states and achieve secure commercial fiber transmission distances of $30.02$~{\rm km} ($12.48$~{\rm dB} loss) and $49.85$~{\rm km} ($11.62$~{\rm dB} loss) in Xi'an and Guangzhou, two cities in China, respectively. The corresponding secret key rates are two orders-of-magnitude higher than that of previous demonstrations~\cite{Huang_OptLett_2016,Jouguet_OptExpr_2012,Fossier_NewJPhys_2009}. We achieve this by developing a fully automatic control system, that achieves stable excess noise, and also by applying a rate-adaptive reconciliation protocol to achieve a high-reconciliation efficiency with high success probability. High secret key rate has been accomplished by using our new proposed one-time calibration model, which provides more efficient calibration procedure as well as simpler experimental implementations.

\begin{figure}[t]
\centerline{\includegraphics[width=12cm]{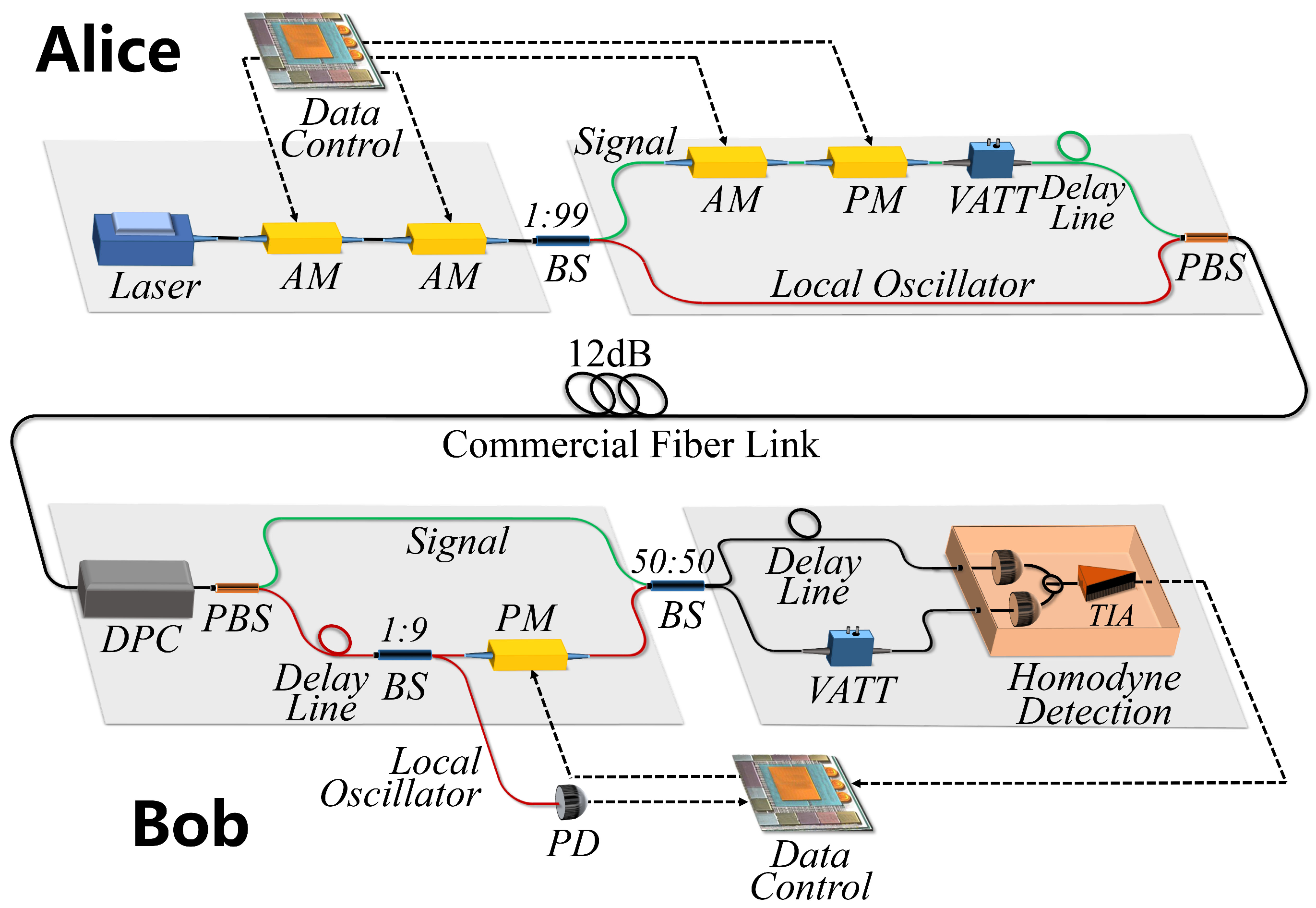}}
\caption{ (Color online) Optical layout for the field tests of the continuous-variable QKD system. Alice sends an ensemble of 40 ns Gaussian-modulated coherent states to Bob with a strong local oscillator multiplexed in time using a delay line and orthogonal polarization via a polarizing beamsplitter. The states are demultiplexed on Bob's side with another polarizing beamsplitter combined with an active dynamic polarization controller.  After demultiplexing, the signal and local oscillator interfere on a shot-noise-limited balanced pulsed homodyne detector with a phase modulator on the local oscillator path to perform the random choice of the measured signal quadrature. Laser: continuous-wave laser; AM: amplitude modulator; PM: phase modulator; BS: beamsplitter; VATT: variable attenuator; PBS: polarizing beamsplitter; DPC: dynamic polarization controller; PD: photodetector.}\label{fig1}
\end{figure}

\section*{Results}

The CV-QKD experimental setup is illustrated in Fig.~1 and consists of two legitimate users, Alice and Bob. To begin with, Alice generates 40ns coherent light pulses by a $1550$~nm telecom laser diode and two high-extinction amplitude modulators. Here the line width of the laser is 0.5 kHz and the extinction ratio of the amplitude modulators is $45$~{\rm dB}. These two amplitude modulators are pulsed with a duty cycle of 20\% with a frequency of 5 MHz. These pulses are split into weak signals and strong local oscillators (LO) with a $1/99$ beamsplitter, which ensures that the LO has enough power. The signal pulse is modulated with a centered Gaussian distribution using an amplitude and a phase modulator, where the modulation data is originally derived from our self-designed quantum random number generator~\cite{Xu_QST_2019, Zheng_QST_2019}. The variance is controlled using a variable attenuator and optimized with the loss of $12$~{\rm dB}. The signal pulse is delayed by $45$~ns with respect to the LO pulse using a $9$~m delay line. The devices used by Alice are polarization-maintaining. Both pulses are multiplexed with an orthogonal polarization using a polarizing beamsplitter. The time and polarization multiplexed pulses are sent to Bob in one fiber to reduce the phase noise caused in the transmission process.

After the fiber link, Bob first uses the dynamic polarization controller to compensate the polarization drift, so that the polarization extinction ratio of polarization demultiplexing is kept at a high level. In our field tests, the extinction ratio of polarization after compensation is about $30$~{\rm dB}. A second delay line on Bob's side, which corresponds to the one used by Alice, allows for a time superposition of both the signal and LO pulses. A part of the LO, around 10\%, is used in the system for clock synchronization, data synchronization, and LO monitor. A phase modulator on the LO path allows for a random selection of the measured signal quadrature and compensation of the phase drift between signal path and LO path. Then the signal and LO interfere on a shot-noise-limited balanced pulsed homodyne detector~\cite{Xiaoxiong_PJ_2018}. A time delay fiber and an attenuator are used to compensate the bias of the beamsplitter and the photodetectors.

\begin{figure*}[t]
\centerline{\includegraphics[width=16cm]{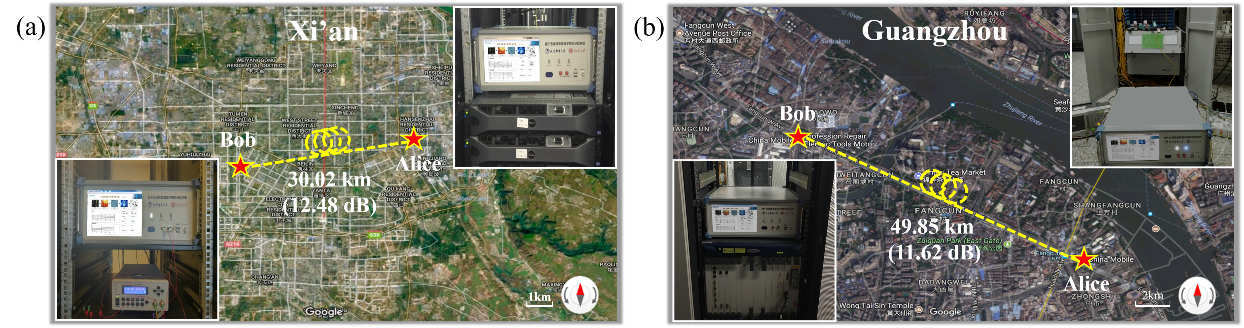}}
\caption{ (Color online) Bird's-eye view of the two field-environment CV-QKD systems. (a) Field test environment in Xi'an, where Alice is situated at a HuoJuLu Service Room (Xi'an Uotocom Network Technology Co., Ltd.) and Bob is situated at a HeShengJingGuang Service Room (Xi'an Uotocom Network Technology Co., Ltd.). The deployed fiber length is $30.02$~{\rm km} ($12.48$~{\rm dB}).  (b) Field test environment in Guangzhou, where Alice is at a QingHeDong Service Room (China Mobile Group Guangdong Co., Ltd.) and Bob is at a FangCun Service Room (China Mobile Group Guangdong Co., Ltd.). The deployed fiber length is $49.85$~{\rm km} ($11.62$~{\rm dB}). \textcircled{c}Google Maps - 2017}\label{fig2}
\end{figure*}

For the fiber link, we undertake the field tests in the commercial fiber networks in two different cities in China. The first one is in Xi'an and operated by Xi'an Uotocom Network Technology Co., Ltd. The second one is in Guangzhou and operated by China Mobile Group Guangdong Co., Ltd. The channel loss of these two fiber links is similar, i.e., approximately $12$~{\rm dB}. However, the transmission distances are very different due to the different types of physical networks. As illustrated in Fig.~2~(a), the first network is operated in the inner city of Xi'an, which has higher channel loss per kilometer. The total deployed fiber length is $30.02$~{\rm km} and the transmission loss is $12.48$~{\rm dB} for the fiber link (0.416 {\rm dB/km}). The second network is between two different districts in Guangzhou, as shown in Fig.~2~(b), the total deployed fiber length is $49.85$~{\rm km} and the transmission loss is $11.62$~{\rm dB} for the fiber link (0.233 {\rm dB/km}). For the first field test, Alice is placed at the site of the HuoJuLu Service Room in Xi'an (\emph{N}$34{^{\circ}}15'12''$, \emph{E}$109^{\circ}0'44''$) and Bob at the site of the HeShengJingGuang Service Room (\emph{N}$34^{\circ}14'8''$, \emph{E}$108^{\circ}53'49''$). For the second field test, Alice is placed at the site of the QingHeDong Service Room in Guangzhou (\emph{N}$22^{\circ}59'20''$, \emph{E}$113^{\circ}24'9''$) and Bob at the site of the FangCun Service Room (\emph{N}$23^{\circ}5'49''$, \emph{E}$113^{\circ}14'8''$).

To overcome the channel perturbations due to changing environmental conditions, we developed several automatic feedback systems to calibrate time, polarization, and phase of the quantum states transmitted. For the time calibration shown in Fig.~1, there are two modules. One is a data synchronization module, and the other one is a clock synchronization module to achieve a synchronous clock in two remote places. Here we utilize the LO to perform the data synchronization as well as the clock synchronization. The traditional way involves using the modulated signal for the data synchronization module, which requires much more data to eliminate the effects of noise (the signal-to-noise ratio (SNR) of the signal is always extremely low ($<1$) when transmitted long distances) and the success possibility cannot reach $100\%$. For the polarization calibration, we adopt a dynamic polarization controller (DPC) comprised of an electric polarization controller (EPC), a polarization beam splitter (PBS), and a photodetector.

With the use of this polarization calibration system operated in real time, the polarization mode is maintained and the power of the LO arm over the signal arm keeps over $30$~{\rm dB}. For the phase calibration, we utilize the phase modulator in the LO path (the same one as the basis sifting) to compensate the phase drift. Alice inserts the reference data in the signal sequence according to a certain period. The period of insertion is determined by the frequency of the phase drift. In the field tests, 100 reference data are inserted into every 1000 quantum signals to perform the phase estimation and compensation since the major contribution of the phase drift comes from the separate arms in Alice and Bob's MZI. The phase drift between the LO and the signal is obtained by the measurement results of the reference signal. Then the feedback voltage of the phase modulator is calculated from the half-wave voltage and is loaded onto the phase modulator in real time.

The shot-noise unit (SNU) can be vital to a CV-QKD system as the calibrated SNU will be treated as a normalization parameter to quantize the quadrature measurement results. According to the security analysis of CV-QKD,  to evaluate the secret key rate we need to first resolve the corresponding covariance matrix, where its elements need to be normalised by the calibrated SNU. Therefore, the calibrated SNU will directly affect the security analysis on secret key rate. In these field tests the quantum signal path in the homodyne detection is randomly cut off to perform the SNU measurement.

After measurement, Alice and Bob will perform postprocessing to distill the final key. The postprocessing of a CV-QKD system contains four parts: basis sifting, parameter estimation, information reconciliation and privacy amplification. To support the high repetition frequency of a CV-QKD system, all parts of the postprocessing need to be executed at high speed. The computational complexity of basis sifting and parameter estimation is low. We can obtain high speed execution on a CPU, where the average speed of basis sifting and parameter estimation can be achieved up to $17.36$~Mbits/s and $16.49$~Mbits/s, respectively.

In the information reconciliation part, we combine multidimensional reconciliation and multi-edge type LDPC codes to achieve high efficiency at low SNRs~\cite{Leverrier_PhysRevA_2008,Richardson_MutiLDPC_2002}. However, the speed of the error correction is limited on a CPU because of the long block code length and many iterations of the belief propagation decoding algorithm. We implement multiple code words  decoding simultaneously based on our GPU and obtain speeds up to $30.39$~Mbits/s on an NVIDIA TITAN Xp GPU~\cite{Xiangyu_SR_2017}. Privacy amplification is implemented by a hash function (Toeplitz matrices in our scheme)~\cite{Krawczyk_Toeplitz_1994,Fung_PhysRevA_2010}. The average speed of privacy amplification can be achieved to 1.35Gbps at any input length on the GPU ~\cite{Xiangyu_PJ_2018}.

Taking finite-size effects into account, the secret key rate, bounded by collective attacks, is given by~\cite{Leverrier_PhysRevA_2010}

\begin{equation}
K = f\left( {1 - \alpha } \right)\left( {1 - FER} \right)\left[ {\beta I\left( {A:B} \right) - \chi \left( {B:E} \right) - \Delta \left( n \right)} \right],
\end{equation}

where$\beta \in[0,1]$ is the reconciliation efficiency, $I(A:B)$ is the classical mutual information between Alice and Bob, $\chi(B:E)$ is the Holevo quantity~\cite{Nielsen_QCQI}, $\Delta \left( n \right)$ is related to the security of the privacy amplification, FER is the frame error rate related to the reconciliation efficiency of a fixed error correction matrix, $f$ is the repetition rate of a QKD system ($5$~{\rm MHz} for our system) and $\alpha$ is the system overhead. The overhead here means the percentage of the signals that cannot be used to distill the final secret keys. In our system, some of the signal will be used to do the phase compensation.

\begin{figure}[t]
\centerline{\includegraphics[width=15cm]{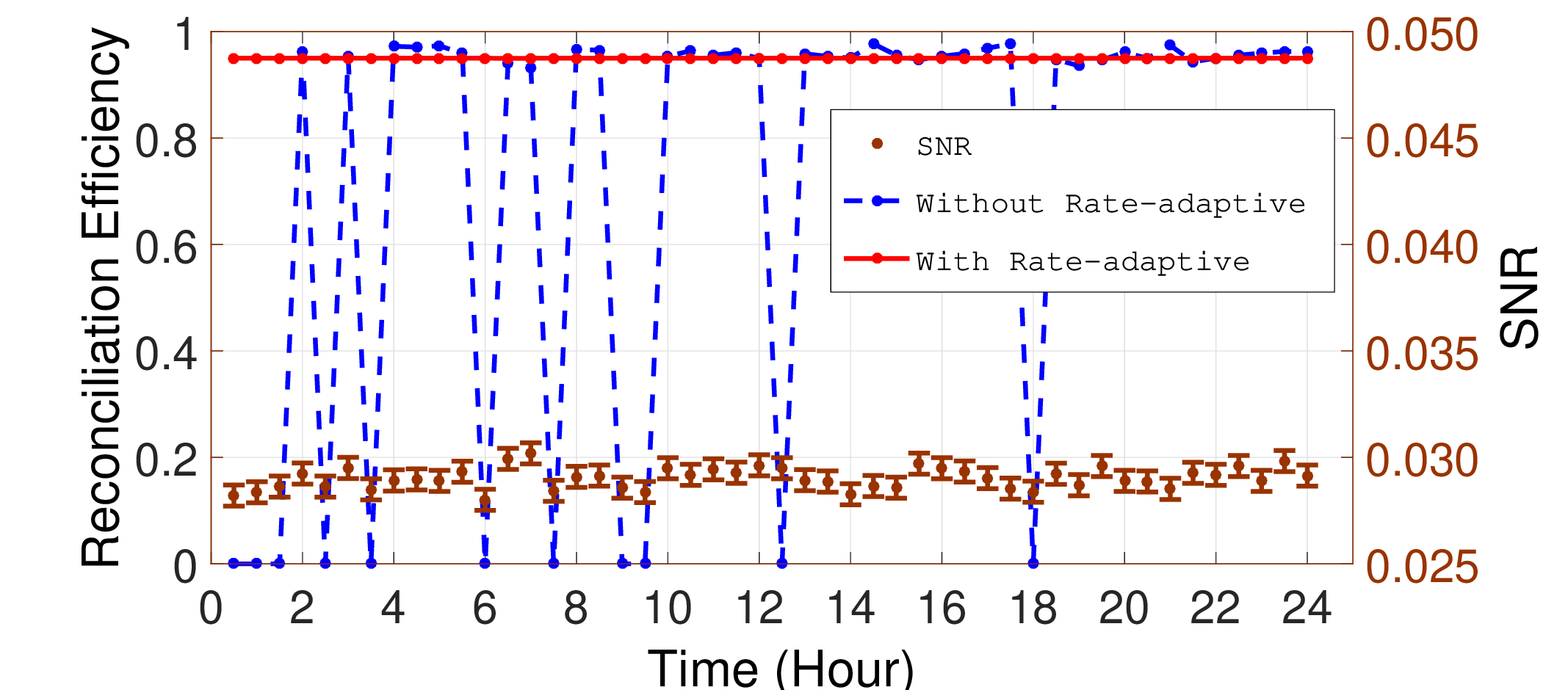}}
\caption{ (Color online) 24 hours continuous test results for the signal-to-noise (SNR) ratio and reconciliation efficiency; in brown, the signal-to-noise-ratio of our system; in blue, the reconciliation efficiency without the rate-adaptive method~\cite{Text}; in red, the reconciliation efficiency with a rate-adaptive method. The reconciliation efficiencies vary greatly without rate-adaptive method which is due to it being very sensitive to the variance of the SNRs of the quantum channel. The reconciliation efficiencies keep a high level with the rate-adaptive method.}\label{fig3}
\end{figure}

To achieve a high key rate, we need to achieve a high reconciliation efficiency and have a low FER, which is a trade off in the experimental realization. For a fixed error correction matrix, achieving a high reconciliation efficiency will lead to a high FER. Thus, in the experiment, we need to find an optimal efficiency and an FER to maximize the secret key rate according to their relationship. Furthermore, we also need to decrease the system overhead to let more pulses distill the final keys.

For the high reconciliation efficiency and low FER part, we utilize the rate-adaptive reconciliation protocol~\cite{Xiangyu_Arxiv_2017}. As shown in Fig.~3, the reconciliation efficiency of decoding with the original MET-LDPC code~\cite{Jouguet_NatPhoton_2013,Xiangyu_Arxiv_2017, Milicevic_NPJ_2018} varies drastically(blue dash dotted line), because the practical SNRs of the quantum channel float in a range. However, rate-adaptive reconciliation~\cite{Xiangyu_Arxiv_2017} can keep high efficiencies even if the SNRs are unstable. For the parity check matrix with a code rate of $0.02$, the maximum reconciliation efficiency of $97.99\%$ can be achieved when the SNR is $0.0287$. Although such high reconciliation efficiency can be achieved, the secret key rate is not necessarily high because the FER is also higher. According to Eq.1, the secret key rate is influenced by both reconciliation efficiency and FER. Technically, the trend of the efficiency and the FER are opposite. Thus, we can find the optimal trade-off between efficiency and FER to maximize the secret key rate. Practically, for the parity check matrix with an original code rate of $0.02$, we can obtain the maximum secret key rate when the practical SNR is $0.0296$. Here the efficiency is $95\%$ and the FER is $0.1$. Thus, we use rate-adaptive reconciliation to maximize the secret key rate of our system by balancing the efficiency and the FER. Otherwise, the secret key rate will be reduced. Actually, it is difficult to find a fixed relationship between FER and reconciliation efficiency. Because it is related to many factors, including the degree distribution and construction methods of error correction codes, the distribution and randomness of the raw keys. The results in the paper are obtained by a fixed error correction code, and this code is chosen in particular.

\begin{figure}[t]
\centerline{\includegraphics[width=12cm]{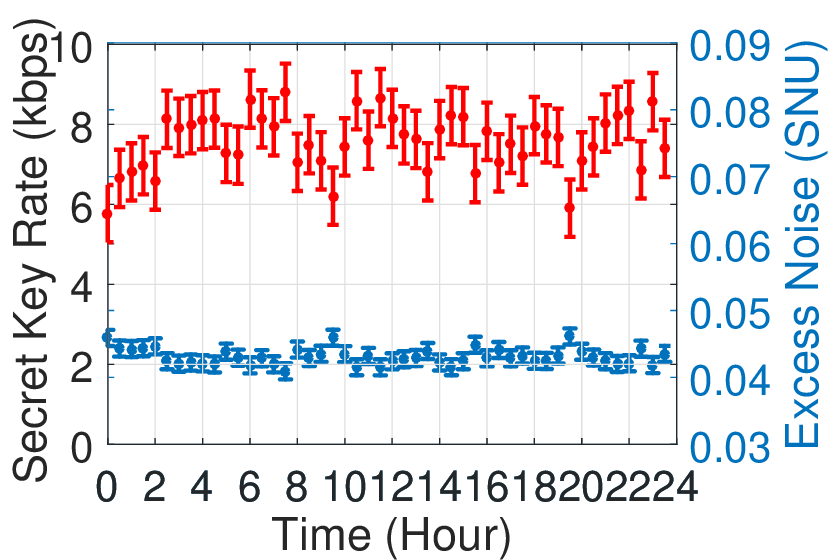}}
\caption{ (Color online) 24 hour continuous test results for the secret key rate and the excess noise; Lower blue mark is the excess noise; Upper red mark represents the the secret key rate.}\label{fig4}
\end{figure}

For the system overhead part, we can swap the order of parameter estimation and information reconciliation to have an almost doubling of the final key rates~\cite{Lupo_PRL_2018, Xiangyu_Arxiv_2018}. According to Eq.1, the ratio of the data which is used to extract the secret keys has an important impact on the secret key rate. In the previous CV-QKD system, there are only $n$ variables used to extract the keys. The other $m = N - n$ variables are used to estimate the quantum channel parameters. Taking into account the influence of the finite-size regime, the ratio of $m$ to $N$ is great for long distance CV-QKD systems. Generally, half of the data is used for parameter estimation, which will reduce the secret key rate by $50\%$. In our system, we swap the order of parameter estimation and information reconciliation. We use all the data to extract the keys, if Alice and Bob successfully correct the errors between them, Alice can recover Bob's raw keys. If the decoding fails, Bob discloses his raw keys. Therefore, no matter the decoding step successes or not, Alice can use the whole raw keys for parameter estimation. But still the failure is unlikely to happen. At the same time, since we use all of the raw keys to distill the secret keys, the secret key rate can be improved.

Using our CV-QKD system integrated with feedback systems, we have accumulated raw data for $24$ hours in  Xi'an and $3$ hours in Guangzhou. During these periods, the automatic feedback systems worked effectively (details in Appendix.A). Compared with laboratory experiments, the field tests faced harsher environmental turbulence.  For instance, the field environment will change the arrival time of the signals. The time calibration system works to monitor the time shift and then compensate it effectively. The achieved timing calibration precision is below 200 ps, which is much smaller than the $40$~{\rm ns} pulse width of the signal laser. Furthermore, with the help of the aforementioned polarization calibration system, we have compensated for the polarization change and achieved a fluctuation less than $5\%$ in the Signal path. Finally, the achieved phase calibration precision is below $1^\circ $ degree, according to the reference~\cite{Huang_SciRep_2015}, the residue phase drift after the phase compensation causes approximately 0.0004 of the excess noise, which is relatively less compare to the total excess noise.

\begin{figure}[t]
\centerline{\includegraphics[width=16cm]{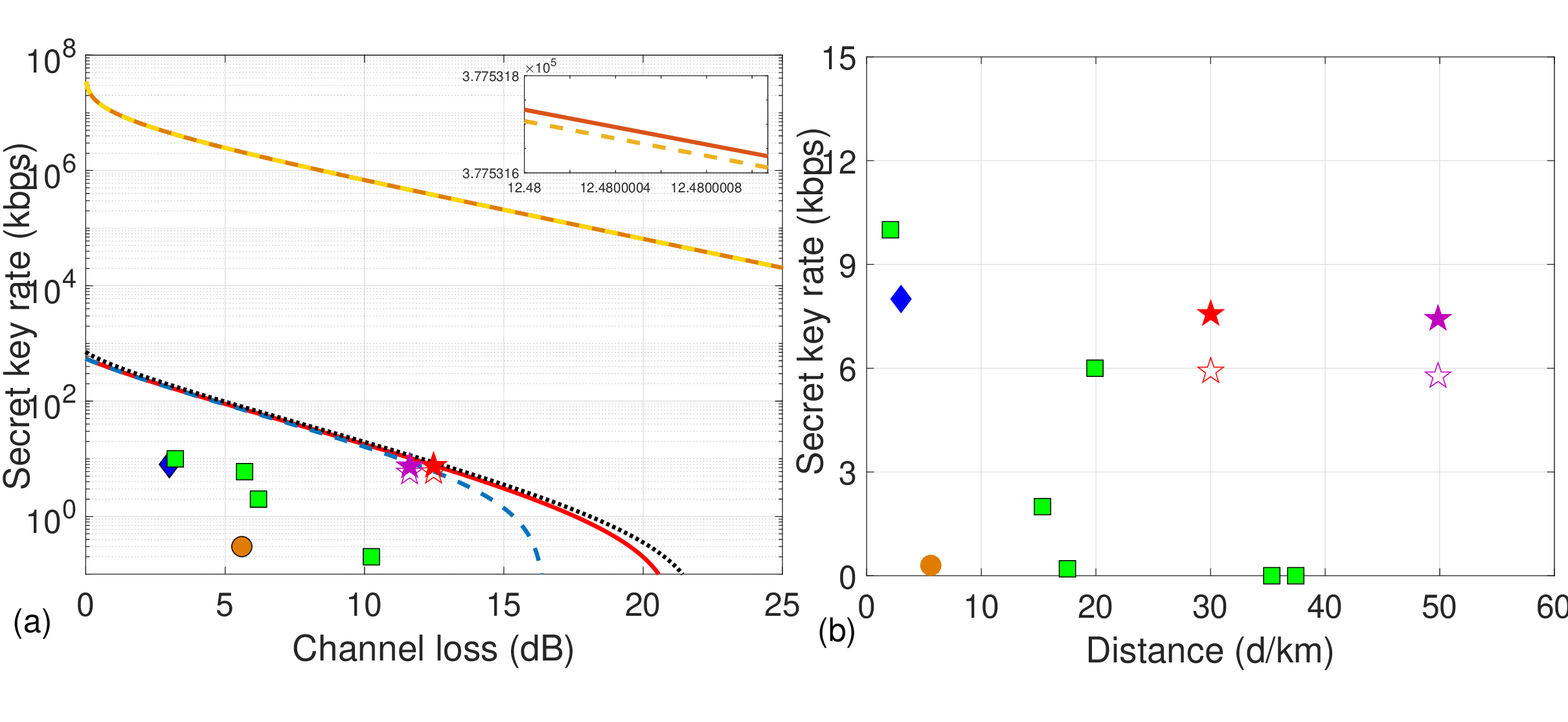}}
\caption{ (Color online) Secure key rates of experiments in the commercial fiber as well as the simulation results shown as a function of fiber loss and length. The red and purple five-pointed stars correspond to the experimental results in Xi'an and Guangzhou with the fiber transmitting losses of $12.48$~{\rm dB} ($30.02$~{\rm km}) and $11.62$~{\rm dB} ($49.85$~{\rm km}), respectively. Solid five-pointed stars are the experimental results under asymptotic limit while the hollow five-pointed stars represent the experimental results under finite-size regime. The black dot line shows the results with the original calibration model under asymptotic limit regime. The red solid curve represents the result calculated with the experimental parameters in the asymptotic limit under one-time-evaluation calibration model. The dashed blue curve represents the result in the finite-size regime with the one-time-evaluation calibration model.
The PLOB bound which provides the fundamental limit of repeaterless quantum communications is also drawn in the figure. This is the brown-yellow line which corresponds to the PLOB bound for a lossy channel~\cite{Pirandola_NatCom_2017} and the one for a thermal-loss channel~\cite{Pirandola_NatCom_2017, Pirandola_QST_2018} considering the average excess noise in our field test. The square points, rhombus point, and the dot point represent the field trials in Ref.~\cite{Huang_OptLett_2016,Jouguet_OptExpr_2012,Fossier_NewJPhys_2009}.}
\end{figure}

To one step further improving the experimental system, we propose and apply a new calibration model with one-time evaluation. Comparing to the rather complicated two-time calibration process, in the one-time evaluation model we only need to measure once the homodyne detector when the LO path is on, and take the result as the shot-noise units (SNU). The new calibration model also attributes to simplify the system implementation as only one optical switch is demanded during the calibration procedure. Moreover, the statistical fluctuation of the SNU introduced by the calibration procedures is reduced. The channel loss versus secret key rate curves have been depicted in Fig.~5, the red solid curve represents the secret key rate in the asymptotic limits when applies our new proposed calibration model, the black dash-dot curve represents the secret key rate in the asymptotic limits but under the original calibration model. It can be observed that the performance of the one-time calibration model and the original two-times calibration model are almost the same except for a slightly different when the channel loss is over 20dB.

The 24 hour continuous test results in the Xi'an network for secret key rates and excess noises are shown in Fig.~4, where the average excess noise of our system is $4\%$ $SNU$.
It should be noticed that the composable security which can help building up a more complete analysis under a practical environment can be carried out in further work. The secret key rate is $7.57$~{\rm kbps} in the asymptotic limit, while the key rate is $5.91$~{\rm kbps} in finite-size regime (details can be found in Appendix.E). The average reconciliation efficiency is 0.9501 with an average SNR of 0.0295 in our continuous test in Guangzhou. The continuous test results in the Guangzhou network for the secret key rate is $7.43$~{\rm kbps} in the asymptotic limit and $5.77$~{\rm kbps} in the finite-size regime (detailed in Appendix.E). Due to the feedback procedure, this CV-QKD system can run continuously for long periods of time automatically.

\section*{Discussion}
With these field tests of a continuous-variable QKD system, we have extended the distribution distance to $50$~{\rm km} over commercial fiber~\cite{Huang_OptLett_2016,Jouguet_OptExpr_2012,Fossier_NewJPhys_2009}. In addition, with an optimized scheme in the optical layer and postprocessing, the secure key rates are higher than previous results by two orders-of-magnitude, which is now comparable to the key rates of discrete-variable QKD systems at metropolitan distances~\cite{Chen_OptExpr_2010,Sasaki_OptExpr_2011,Stucki_NewJPhys_2011,Wang_OptExpr_2014,Shimizu_JLightwaveTechnol_2014}. The PLOB bounds which show the maximum achievable secret key rate in QKD have also been drawn in Fig.5 for both the lossy channel and thermal-loss channel. Although we have achieved highest secret key rates among the reported field tests, the secret key rates are still lower than the PLOB bounds for both the lossy channel and thermal-loss channel.
The secret key rate we achieved here is based on the system with a repetition rate which is only 5~{\rm MHz}. This repetition rate which is much smaller comparing to discrete-variable QKD systems (around $1$~{\rm GHz}). These results have moved continuous-variable QKD towards a more practical setting and the expectation that a secure metropolitan network could be built and is within reach of current technology.

\section*{Appendix}

\subsection*{Appendix A: Experimental details}
There are three automatic feedback systems to calibrate time, polarization, phase of the quantum states transmitted in our system in detail. These three feedback systems are the key modules to run a CV-QKD system in deployed commercial fibers.

For the time calibration shown in Fig.1, there are two modules. One is data synchronization module, and another one is clock synchronization module to get synchronous clock in two remote places. For the data synchronization module, here we utilize a frame synchronization module based on the modulation of the LO. Previously, the traditional way is using the modulated signal, which requires much more data to eliminate the effects of noise (The signal to noise ratio (SNR) of Signal is always extremely low ($<1$) when transmitted long distance) and the success possibility can not reach $100\%$. We modulate data synchronization information on the LO, whose SNR is high enough to reduce the time of synchronization and promote the success possibility to $100\%$. The data synchronization is used to decide the starting point of the signal, so that the receiving system starts to work. It is the premise of the normal operation of the system. The data synchronization signal is a specific train sequence and modulated on the first high-extinction AM for the pulse generation. Because the clock is precisely synchronized in our system, the data synchronization is implemented with a low frequency (0.5kHz) which introduce ignorable effect on shot-noise calibration and the key distribution. For the clock synchronization module, we also utilize LO to accomplish that. Bob utilizes a photoelectric detector (PD) to detect the part of LO pulses. The output signals of the PD is shaped by the comparator, the shaped pulses (5 MHz) are inputted into a clock chip to generate the high frequency clock (200 MHz) which is required by the other modules of the system. In the previous implementation~\cite{Jouguet_NatPhoton_2013}, there is only the realization of clock synchronization by LO but no mention of the data synchronization.

For the polarization calibration, we adopt a polarization stabilization system comprised of an electric polarization controller (EPC), a polarization beam splitter (PBS), and a photodetector. We insert the EPC and PBS after the fiber link, where the transmission port of PBS is connected to Lo path and the reflection port of the PBS is connected to Signal path. One percent of the power of LO pluses are utilized to monitor the polarization state and calculate the feedback voltage of EPC. With the use of this polarization calibration system operated in real time, the polarization mode is maintained and the power ratio of the LO and the Signal keeps 30 dB.  The all-fiber construction of the calibration system provides very low insertion loss. Compared to the polarization compensation in Ref.[5], distilling the drift information needs more homodyne detection results than the scheme based on detecting the leaked LO in signal path. When the repetition frequency of the system is higher than polarization drift ratio, the amount of detection results is large enough to calculate drift. But in order to deal with various situations where polarization drift may be severe, the polarization controller based on the high-SNR LO pulses is more applicable.

For the phase calibration, although the LO and the Signal is transmitted in one fiber with a small delay, a slow phase shift which is aroused from the difference between the LO path and Signal path is unavoidable. To stabilize this, we utilize the phase modulator in the LO path (the same one as the base sifting) to compensate the phase drift. Alice inserts the reference data in the signal sequence according to a certain period. The period of insertion is determined by the frequency of the phase drift. The phase drift between the LO and the Signal is obtained by the measurement results of the reference signal. Then the feedback voltage of the phase modulator is calculated from the half-wave voltage and is loaded onto the phase modulator in real time.

\subsection*{Appendix B: Calibration model with one-time evaluation.}

In the previous experimental demonstrations, two-times calibration process is extensively used as it requires firstly measuring the homodyne detector output when both LO path and signal path are off, then the homodyne detector output when only LO path is connected. Finally using subtraction to calculate the SNU. In the proposed one-time-evaluation (OTE) calibration model, we only need to measure once the homodyne detector when the LO path is on, thus the SNU of these two models can be deduced as:
\begin{equation}
\begin{array}{l}
SN{U^{TTE}} = {V_{tot}} - {V_{ele}},\\
SN{U^{OTE}} = {V_{tot}} = SN{U^{TTE}} + {V_{ele}}.
\end{array}
\end{equation}
Where $V_{tot}$ and $V_{ele}$ are corresponding to the variance of the output of the homodyne detector when the LO path is on and off.  ${SN{U^{TTE}}}$ represents the two-times calibration SNU and ${SN{U^{OTE}}}$ represents the SNU with one-time calibration.

Under this modeling, certain advantages can be procured: Firstly, we only need one optical switch in signal path in our system. Secondly, only one-time statistic evaluation is required, which makes it a more utility model applying in practical systems. Thirdly, since we only need one-time period to calculate the SNU, the statistic fluctuation is minimized compared to original calibration model with two-times evaluation.

We now give a more detailed analysis on its availability. We first consider the output of a practical homodyne detector with limited detection efficiency and electronic noise:
\begin{equation}
{X_{out}} = A{X_{LO}}\left( {\sqrt {{\eta _d}} {{\hat x}_B} + \sqrt {1 - {\eta _d}} {{\hat x}_{v1}}} \right) + {X_{ele}}.
\end{equation}
Where ${\hat x_B}$ represents the x-quadrature of the canonical components of mode after the transmission and ${\hat x_{{v_1}}}$ represents the vacuum state.  ${X_{ele}}$ is a Gaussian variable with variance ${v_{el}}$, ${A}$ is the circuit amplification parameter.
The output of the homodyne detector need to be further quantized by the SNU to estimate Eve's information. In two-times calibration model, it is $SNU^{TTE} = {A^2}X_{LO}^2$. In one-time-evaluation model, it is ${SNU^{OTE} = {A^2}X_{LO}^2 + \left\langle {X_{ele}^2} \right\rangle  = {A^2}X_{LO}^2 + {v_{el}}.}$ Then the data used for postprocessing is:
\begin{equation}
x_{out}^{OTE} = \frac{{A{X_{LO}}}}{{\sqrt {{A^2}X_{LO}^2 + {v_{el}}} }}\left( {\sqrt {{\eta _d}} {{\hat x}_B} + \sqrt {1 - {\eta _d}} {{\hat x}_{v_1}}} \right) + \frac{{\sqrt {{v_{el}}} }}{{\sqrt {{A^2}X_{LO}^2 + {v_{el}}} }}{\hat x_{v_2}}.
\end{equation}

Here the Gaussian variable ${X_{ele}}$ in Eq.(3) is replaced with a Gaussian operator $\sqrt {{v_{el}}} {\hat x_{{v_2}}}$ since the electronic noise is not controlled by Eve, ${v_{el}}$ is the variance of ${X_{ele}}$ and ${\hat x_{v_2}}$ is the vacuum of variance 1.
If let ${{\eta _e} = \frac{{{A^2}X_{LO}^2}}{{{A^2}X_{LO}^2 + {v_{el}}}}}$, it can adequately characterize the electronic noise using a beamsplitter.

The completely version of the EB model is depicted in the upper side of Fig.~6. Alice prepares the EPR state that has two modes A and B where mode A is kept in Alice's side and will be measured by heterodyne detection. Mode B will be sent into the channel, in which Eve can conduct any attack according to her strategies. After the transmission, mode B will pass through two beamsplitters and become mode B' before the ideal homodyne detection.
Based on the trusted model assumption, the limited detcted efficiency and electronic noise is not controlled by Eve thus if we commute the order of the beamsplitters it will not change the final detected mode B', the EB model after the beamsplitter switching is depicted on the bottom of Fig.~6.
Under this modelling, we no longer need to measure the electronic noise.

\begin{figure*}[t]
\centerline{\includegraphics[width=15cm]{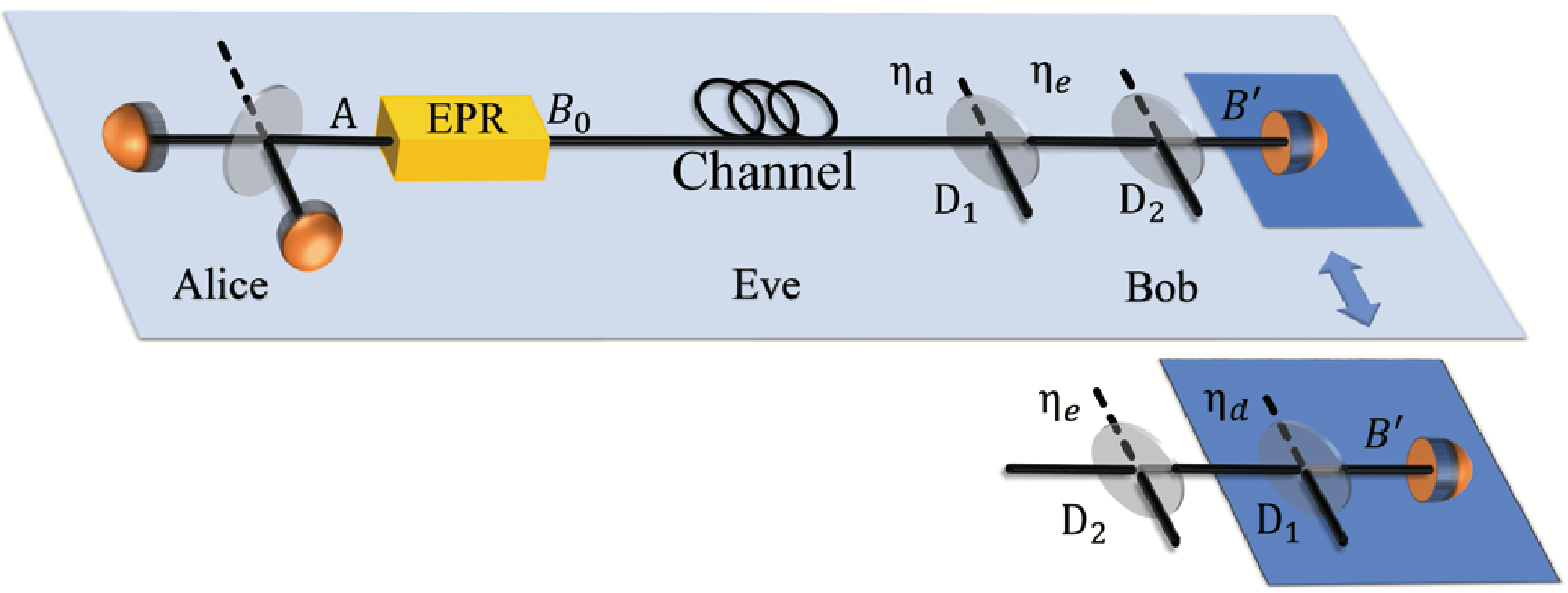}}
\caption{ (Color online) Entanglement-based model of one-time calibration model. Two beamsplitters are brought in to imitate the limited detection efficiency and the detector electronic noise. EPR states correspond to the two mode squeezed states that has two modes A and B, mode A is measured by Alice using heterodyne detection and mode B is sent to Bob. The mode arrived before the ideal homodyne detection is B'.}
\end{figure*}

\subsection*{Appendix C: Secret key rate with one-time-evaluation calibration model.}

The (asymptotical) secret key rate ${K}$ with one-time-evaluation calibration model against collective attacks for reverse reconciliation is given by~\cite{Devetak_ProcRSoc_2005}

\begin{equation}
K = f(1 - \alpha )(1 - FER)[\beta I(A:B) - \chi (B:E)].
\end{equation}

The classical information of Alice and Bob $I(A:B)$ can be described by Shannon entropy which can be calculated by the variance of Alice and Bob and their covariance:
\begin{equation}
I\left( {A:B} \right) = \frac{1}{2}{\log _2}(\frac{{{V} + \chi }}{{\chi  + 1}}).
\end{equation}
Here we define ${\chi}$ as ${\chi  = \frac{1}{{T{\eta _d}{\eta _e}}} - 1 + {\varepsilon _c}}$ for conciseness, ${T}$ is the channel transmittance and ${\varepsilon _c}$ stands for the channel excess noise.

Applying one-time-evaluation model, we no longer need to measure the electronic noise, so mode ${D_2}$ in the Fig.~
5 is unknown to us.
To simplify our calculation, however, the co-variance matrix of mode ${A, D_1, B'}$ can be easily obtained:
\begin{equation}
{\gamma _{A{D_1}{B'}}} = \left( {\begin{array}{*{20}{c}}
{{\gamma _A}}&{{\phi _{A{D_1}}}}&{{\phi _{A{B'}}}}\\
{\phi _{A{D_1}}^T}&{{\gamma _{D_1}}}&{{\phi _{{D_1}{B'}}}}\\
{\phi _{A{B'}}^T}&{\phi _{{D_1}{B'}}^T}&{{\gamma _{{B'}}}}
\end{array}} \right).
\end{equation}
Thus we write the Holevo quantity~\cite{Nielsen_QCQI} ${\chi \left( {B:E} \right)}$ which restricts the upper bound of the information that the eavesdropper Eve can acquire as:
\begin{equation}
\chi (B:E) = \chi (B:E,{D_2}) = S({\rho _{A{D_1}B'}}) - S(\rho _{A{D_1}}^{{m_{B'}}}).
\end{equation}

The first term of the right side of the equation can be calculated by its corresponding co-variance matrix ${{\gamma _{A{D_1}B'}}}$ using its symplectic eigenvalues, the covariance matrix is derived as:
\begin{equation}
\scriptsize{
{\gamma _{A{B'}{D_1}}} = \left( {\begin{array}{*{20}{c}}
{V{I_2}}&{\sqrt {T{\eta _e}{\eta _d}({V^2} - 1)} {\sigma _z}}&{\sqrt {T{\eta _e}(1 - {\eta _d})({V^2} - 1)} {\sigma _z}}\\
{\sqrt {T{\eta _e}{\eta _d}({V^2} - 1)} {\sigma _z}}&{[T{\eta _e}{\eta _d}(V - 1 + {\varepsilon _c}) + 1]{I_2}}&{\sqrt {{\eta _d}(1 - {\eta _d})} T{\eta _e}(V - 1 + {\varepsilon _c}){I_2}}\\
{\sqrt {T{\eta _e}(1 - {\eta _d})({V^2} - 1)} {\sigma _z}}&{\sqrt {{\eta _d}(1 - {\eta _d})} T{\eta _e}(V - 1 + {\varepsilon _c}){I_2}}&{[T{\eta _e}(1 - {\eta _d})(V - 1 + {\varepsilon _c}) + 1]{I_2}}
\end{array}} \right).
}
\end{equation}

Experimentally, using Alice's data and Bob's data is sufficiently to obtain the parameter values appeared in the co-variance matrix above. The Von Neumann entropy ${S(\rho _{AC}^{{m_{B'}}})}$ is calculated from matrix ${\gamma _{A{D_1}}^{{m_{B'}}}}$, which is the matrix after Bob performs homodyne detection on mode ${B'}$, it can be derived as:
\begin{equation}
\gamma _{A{D_1}}^{{m_{B'}}} = {\gamma _{A{D_1}}} - {\phi _{A{D_1}B'}}{(X{\gamma _{B'}}X)^{MP}}\phi _{A{D_1}B'}^T.
\end{equation}
From co-variance matrix ${{\gamma _{A{B'}{D_1}}}}$ we can find three valid symplectic eigenvalues and from ${\gamma _{A{D_1}}^{{m_{B'}}}}$ we can find two valid symplectic eigenvalues. The ${\chi (B:E)}$ can now rewritten as:
\begin{equation}
 \chi (B:E) = \sum\nolimits_{i = 1}^3 {G(\frac{{{\lambda _i} - 1}}{2}) - } \sum\nolimits_{i = 4}^5 {G(\frac{{{\lambda _i} - 1}}{2})} .
\end{equation}
Where ${G(x) = (x + 1){\log _2}(x + 1) - x{\log _2}x.}$ Then the final secret key rate is obtainable by using the equation (5). It should be noted that this calculation will lead to a secure lower bound of the actual secret key rate.

\subsection*{Appendix D: Rate-adaptive reconciliation protocol.}

The rate-adaptive reconciliation protocol is to maximize the secret key rate of our system by balancing the efficiency $\beta$ and FER. With the combination of puncturing and shortening techniques, the code rate will be changed to
\noindent
\begin{equation}
R=\frac{n-m-s}{n-p-s}
\,, \label{coderate}
\end{equation}
where the original code rate is $R^{o}=(n-m)/n$, $n$ is the length of the code and $m$ is the length of redundancy bits. The code rate will be changed to $R^{'}=(n-m)/(n-p)$ by adding punctured bits of length $p$. The code rate will be changed to $R^{''}=(n-m-s)/(n-s)$ by adding shortened bits of length $s$.

Therefore, we could have the reconciliation efficiency, which is defined as follows:
\noindent
\begin{equation}
\beta=\frac{R}{C}
\,, \label{Reconciliationefficiency}
\end{equation}
where $R$ is the rate of MET-LDPC code, $C$ is the classical capacity of the quantum channel, which is $C=\frac{1}{2}log_{2}(1+SNR)$ for Gaussian variables.

The detailed steps of the rate-adaptive protocol are as follows~\cite{Xiangyu_Arxiv_2017}:

{\it Step~1}: According to the practical SNR of the experimental data ($0.028$ ~ $0.030$), we calculate the optimal code rate. Then we select a good-performance original MET-LDPC code $0.02$ whose code rate is close to the optimal code rate.

{\it Step~2}: We creates a new sequence $\widehat{u}$ with length $n$ in Bob's side:
\begin{equation}
u=\{u_{1},u_{2},\cdots\cdots,u_{n-p-s}\}\,,
\end{equation}
\begin{equation}
p_{B}=\{{p_{B}}_{1},{p_{B}}_{2},\cdots\cdots,{p_{B}}_{p}\}\,,
\end{equation}
\begin{equation}
s_{B}=\{{s_{B}}_{1},{s_{B}}_{2},\cdots\cdots,{s_{B}}_{s}\}\,,
\end{equation}
\begin{equation}
\widehat{u}=\{u_{1},u_{2},\cdots,{s_{B}}_{1},\cdots,{p_{B}}_{1},\cdots,{s_{B}}_{i},\cdots,{p_{B}}_{j},\cdots,u_{k},\cdots\}\,,
\end{equation}
where $u$ is the string for Bob's multidimensional reconciliation with length $n-p-s$, the sequences $p_{B}$ and $s_{B}$ represent Bob's punctured bits and shortened bits which are randomly generated by Bob with length $p$ and $s$, and $i\in{\{1,2,\cdots,s\}}$, $j\in{\{1,2,\cdots,p\}}$, $k\in{\{1,2,\cdots,n-p-s\}}$. The new string $\widehat{u}$ is created by randomly inserting $p_{B}$ and $s_{B}$ into the string $u$. Then Bob calculates the syndrome of $\widehat{u}$, such that $c(\widehat{u})=H\widehat{u}^{T}$.
$H$ matrix is the parity check matrix of the code, which is used to assist in the encoding and decoding process. Its rows refer to the check nodes and columns refer to the bit nodes. The nodes are obtained by the degree distribution. And H matrix is obtained by some construction method according to the degree distribution.

{\it Step~3}: After we receives the message sent by Bob in Alice's side, we constructs a new sequence $\widehat{v}$ with length $n$:
\begin{equation}
v=\{v_{1},v_{2},\cdots\cdots,v_{n-p-s}\}\,,
\end{equation}
\begin{equation}
p_{A}=\{{p_{A}}_{1},{p_{A}}_{2},\cdots\cdots,{p_{A}}_{p}\}\,,
\end{equation}
\begin{equation}
\widehat{v}=\{v_{1},v_{2},\cdots,{s_{B}}_{1},\cdots,{p_{A}}_{1},\cdots,{s_{B}}_{i},\cdots,{p_{A}}_{j},\cdots,v_{k},\cdots\}\,,
\end{equation}
where $v$ is the sequence for Alice's multidimensional reconciliation with length $n-p-s$, the sequence $p_{A}$ represents Alice's punctured bits which is randomly generated by Alice, and $i\in{\{1,2,\cdots,s\}}$, $j\in{\{1,2,\cdots,p\}}$, $k\in{\{1,2,\cdots,n-p-s\}}$. The new string $\widehat{v}$ and $\widehat{u}$ have the same positions and length of punctured bits and shortened bits. And they also have the same value of shortened bits $s_{B}$. Then Alice uses the belief propagation decoding algorithm or some other algorithm to recover $\widehat{u}$. Finally Alice and Bob will share a common string.

\subsection*{Appendix E: The detailed results of the field test in Xi'an and Guangzhou commercial fiber networks}

~\\
~\\
~\\
~\\

\begin{table}[hbtp]
\scriptsize{
\centering
\caption{The detailed results of the field test in Xi'an commercial fiber network for 24 hours. The detection efficiency of the homodyne detector is 0.612. SNR: practical signal-to-noise ratio. $\beta^{o}$: reconciliation efficiency of the original code. $\beta$: reconciliation efficiency of the rate-adaptive reconciliation. s: the length of shortened bits. p: the length of punctured bits. $\epsilon$: excess noise. ${\eta _e}$: electronic noise $k_{Asymptotic}$: final secret key rate in asymptotic limits. $k_{finite}$: finial secret key rate in finite-size regime.}
\begin{tabular}{|p{1.0cm}<{\centering}||p{1.2cm}<{\centering}|p{1.2cm}<{\centering}|p{1.2cm}<{\centering}|p{1.0cm}<{\centering}|p{1.0cm}<{\centering}|p{1.8cm}<{\centering}|p{1.8cm}<{\centering}|p{1.8cm}<{\centering}|p{1.8cm}<{\centering}|}
  \hline

Time	&	SNR 	&	$\beta^{o}$	&  $\beta$      &	s	&	p	&	$\epsilon$	& ${{v_{el}}}$ & 	$k_{Asymptotic}$	(kbps) &	$k_{finite}$	(kbps) \\\hline
0:30	&	0.028205	&	0	&	0.950268	&	952		&	0		&	0.045909	&	0.257087972	&	5.768156006	&	4.1097321	\\\hline
1:00	&	0.028365	&	0	&	0.950043	&	848		&	0		&	0.044316	&	0.24805773	&	6.647619698	&	4.989195793	\\\hline
1:30	&	0.028628	&	0	&	0.950305	&	664		&	0		&	0.044234	&	0.212818655	&	6.811478575	&	5.153054669	\\\hline
2:00	&	0.029241	&	0.961982	&	0.950289	&	248		&	0		&	0.044372	&	0.218014447	&	6.966862114	&	5.308438208	\\\hline
2:30	&	0.028632	&	0	&	0.950174	&	664		&	0		&	0.044643	&	0.229307098	&	6.583656263	&	4.925232357	\\\hline
3:00	&	0.029504	&	0.953529	&	0.950165	&	72		&	0		&	0.042427	&	0.221973702	&	8.126644238	&	6.468220332	\\\hline
3:30	&	0.028497	&	0	&	0.950352	&	752		&	0		&	0.042061	&	0.136639115	&	7.927945618	&	6.269521712	\\\hline
4:00	&	0.028918	&	0.972573	&	0.950069	&	472		&	0		&	0.042218	&	0.139557273	&	7.992647763	&	6.334223858	\\\hline
4:30	&	0.028968	&	0.970918	&	0.950357	&	432		&	0		&	0.042114	&	0.136224858	&	8.093704911	&	6.435281006	\\\hline
5:00	&	0.028893	&	0.973403	&	0.950115	&	488		&	0		&	0.04195	&	0.139011213	&	8.131799003	&	6.473375098	\\\hline
5:30	&	0.029327	&	0.959201	&	0.950175	&	192		&	0		&	0.043854	&	0.041323934	&	7.272916672	&	5.614492767	\\\hline
6:00	&	0.028012	&	0	&	0.950024	&	1088		&	0		&	0.042957	&	0.040551421	&	7.229510101	&	5.571086196	\\\hline
6:30	&	0.02993	&	0.940153	&	0.949994	&	0		&	10368		&	0.04181	&	0.052974806	&	8.63214663	&	6.973722724	\\\hline
7:00	&	0.030191	&	0.932144	&	0.949996	&	0		&	10360		&	0.042883	&	0.05196541	&	8.137967712	&	6.479543807	\\\hline
7:30	&	0.028433	&	0	&	0.950132	&	800		&	0		&	0.041958	&	0.11270251	&	7.938746112	&	6.280322206	\\\hline
8:00	&	0.029086	&	0.967035	&	0.95035	&	352		&	0		&	0.040907	&	0.11184432	&	8.80254033	&	7.144116424	\\\hline
8:30	&	0.029145	&	0.965105	&	0.950346	&	312		&	0		&	0.044158	&	0.142472731	&	7.051915692	&	5.393491787	\\\hline
9:00	&	0.028574	&	0	&	0.950145	&	704		&	0		&	0.042901	&	0.148243261	&	7.491192309	&	5.832768403	\\\hline
9:30	&	0.028368	&	0	&	0.950333	&	840		&	0		&	0.043532	&	0.110357487	&	7.08897527	&	5.430551365	\\\hline
10:00	&	0.02949	&	0.953975	&	0.950235	&	80		&	0		&	0.04593	&	0.109311065	&	6.204816116	&	4.54639221	\\\hline
10:30	&	0.02918	&	0.963964	&	0.950357	&	288		&	0		&	0.043481	&	0.159636778	&	7.434540248	&	5.776116342	\\\hline
11:00	&	0.029437	&	0.955668	&	0.950048	&	120		&	0		&	0.041518	&	0.154011192	&	8.590946155	&	6.932522249	\\\hline
11:30	&	0.02928	&	0.960719	&	0.950171	&	224		&	0		&	0.043213	&	0.156899036	&	7.604695583	&	5.946271677	\\\hline
12:00	&	0.029627	&	0.949627	&	0.95	&	0		&	392		&	0.041517	&	0.160396589	&	8.668259572	&	7.009835666	\\\hline
12:30	&	0.02819	&	0	&	0.950375	&	960		&	0		&	0.042405	&	0.145677068	&	8.150250582	&	6.491826677	\\\hline
13:00	&	0.02935	&	0.95846	&	0.950193	&	176		&	0		&	0.042722	&	0.154258455	&	7.735690531	&	6.077266626	\\\hline
13:30	&	0.029502	&	0.953593	&	0.950228	&	72		&	0		&	0.042887	&	0.146898664	&	7.62197071	&	5.963546804	\\\hline
14:00	&	0.0296	&	0.950481	&	0.950108	&	8		&	0		&	0.043946	&	0.147434369	&	6.813599789	&	5.155175883	\\\hline
14:30	&	0.028777	&	0.977271	&	0.950056	&	568		&	0		&	0.042241	&	0.204325891	&	7.871299729	&	6.212875823	\\\hline
15:00	&	0.029437	&	0.955668	&	0.950048	&	120		&	0		&	0.041536	&	0.199646968	&	8.219492648	&	6.561068742	\\\hline
15:30	&	0.029704	&	0.947201	&	0.949998	&	0		&	2944		&	0.042426	&	0.168286976	&	8.19474906	&	6.536325154	\\\hline
16:00	&	0.029493	&	0.953879	&	0.95014	&	80		&	0		&	0.044886	&	0.178200272	&	6.767835413	&	5.109411508	\\\hline
16:30	&	0.029338	&	0.958846	&	0.9502	&	184		&	0		&	0.042852	&	0.193986108	&	7.827893557	&	6.169469651	\\\hline
17:00	&	0.029028	&	0.96894	&	0.950321	&	392		&	0		&	0.044096	&	0.19820625	&	7.038879834	&	5.380455928	\\\hline
17:30	&	0.028786	&	0.97697	&	0.950147	&	560		&	0		&	0.042854	&	0.187691506	&	7.501930063	&	5.843506157	\\\hline
18:00	&	0.028316	&	0	&	0.950104	&	880		&	0		&	0.043287	&	0.188699394	&	7.208032562	&	5.549608656	\\\hline
18:30	&	0.029687	&	0.947736	&	0.949993	&	0		&	2376		&	0.042483	&	0.112187631	&	7.96854483	&	6.310120924	\\\hline
19:00	&	0.03007	&	0.935839	&	0.949998	&	0		&	14904		&	0.042464	&	0.10987308	&	7.762217726	&	6.103793821	\\\hline
19:30	&	0.029721	&	0.946667	&	0.949996	&	0		&	3504		&	0.043288	&	0.321548662	&	7.672138455	&	6.013714549	\\\hline
20:00	&	0.029262	&	0.961301	&	0.950371	&	232		&	0		&	0.046126	&	0.338690481	&	5.903249913	&	4.244826007	\\\hline
20:30	&	0.029622	&	0.949785	&	0.949998	&	0		&	224		&	0.043842	&	0.144982226	&	7.08805103	&	5.429627124	\\\hline
21:00	&	0.028868	&	0.974234	&	0.950162	&	504		&	0		&	0.042965	&	0.143486744	&	7.437802729	&	5.779378823	\\\hline
21:30	&	0.028632	&	0	&	0.95017	&	664		&	0		&	0.044643	&	0.207438908	&	8.033042141	&	6.374618236	\\\hline
22:00	&	0.029504	&	0.953529	&	0.95016	&	72		&	0		&	0.042427	&	0.204039996	&	8.22393663	&	6.565512724	\\\hline
22:30	&	0.028497	&	0	&	0.95035	&	752		&	0		&	0.042061	&	0.321594474	&	8.349520597	&	6.691096691	\\\hline
23:00	&	0.028918	&	0.972573	&	0.95007	&	472		&	0		&	0.042218	&	0.338761392	&	6.861580907	&	5.203157002	\\\hline
23:30	&	0.02983	&	0.943258	&	0.949999	&	0		&	7096		&	0.041853	&	0.256120393	&	8.56634362	&	6.907919714	\\\hline
0:00	&	0.029605	&	0.950323	&	0.950323	&	0		&	0		&	0.043515	&	0.261181193	&	7.399916031	&	5.741492125	\\\hline

\end{tabular}
}
\end{table}

\begin{table}[h]
\scriptsize{
\centering
\caption{The detailed results of the field test in Guangzhou commercial fiber network for continuous 48 blocks. The detection efficiency of the homodyne detector is 0.612. SNR: practical signal-to-noise ratio. s: the length of shortened bits. p: the length of punctured bits. $\beta$: reconciliation efficiency of the rate-adaptive reconciliation. $\beta^{o}$: reconciliation efficiency of the original code. ${\eta _e}$: electronic noise. $\epsilon$: excess noise. $k_{Asymptotic}$: final secret key rate in asymptotic limits. $k_{finite}$: finial secret key rate in finite-size regime.}
\begin{tabular}{|p{1.0cm}<{\centering}||p{1.2cm}<{\centering}|p{1.2cm}<{\centering}|p{1.2cm}<{\centering}|p{1.0cm}<{\centering}|p{1.0cm}<{\centering}|p{1.8cm}<{\centering}|p{1.8cm}<{\centering}|p{1.8cm}<{\centering}|p{1.8cm}<{\centering}|}
  \hline

Time	&	SNR 	&	$\beta^{o}$	&  $\beta$      &	s	&	p	&	$\epsilon$	& ${{v_{el}}}$ & 	$k_{Asymptotic}$	(kbps) &	$k_{finite}$	(kbps) \\\hline

1	&	0.030241	&	0.930625	&	0.949997	&	0	&	20392	&		0.042136	&	0.392581912	&	8.579503464	&	6.921079558	\\\hline
2	&	0.030869	&	0.911972	&	0.949994	&	0	&	40024	&		0.041306	&	0.36425003	&	9.320395083	&	7.661971177	\\\hline
3	&	0.030093	&	0.935135	&	0.949992	&	0	&	15640	&		0.042493	&	0.399424436	&	8.316653184	&	6.658229278	\\\hline
4	&	0.028783	&	0.97707	&	0.950245	&	560	&	0	&		0.044921	&	0.463095576	&	6.496388751	&	4.837964845	\\\hline
5	&	0.030063	&	0.936054	&	0.949992	&	0	&	14672	&		0.042497	&	0.400822547	&	8.302073854	&	6.643649948	\\\hline
6	&	0.029988	&	0.938361	&	0.949996	&	0	&	12248	&		0.042856	&	0.404315349	&	8.070705358	&	6.412281452	\\\hline
7	&	0.030522	&	0.922184	&	0.95	&	0	&	29280	&		0.042205	&	0.379741876	&	8.658056183	&	6.999632277	\\\hline
8	&	0.030833	&	0.91302	&	0.949994	&	0	&	38920	&		0.042003	&	0.365815545	&	8.903768839	&	7.245344933	\\\hline
9	&	0.030322	&	0.928176	&	0.949995	&	0	&	22968	&		0.042875	&	0.388825962	&	8.195404522	&	6.536980616	\\\hline
10	&	0.029675	&	0.948113	&	0.949998	&	0	&	1984	&		0.043563	&	0.419116752	&	7.553054985	&	5.894631079	\\\hline
11	&	0.029938	&	0.939905	&	0.949998	&	0	&	10624	&		0.04319	&	0.406649656	&	7.864232652	&	6.205808746	\\\hline
12	&	0.029998	&	0.938053	&	0.95	&	0	&	12576	&		0.043012	&	0.403840032	&	7.98785762	&	6.329433714	\\\hline
13	&	0.028429	&	0	&	0.950264	&	800	&	0	&		0.045778	&	0.481301599	&	5.914846678	&	4.256422772	\\\hline
14	&	0.028628	&	0	&	0.950305	&	664	&	0	&		0.045419	&	0.471006441	&	6.179197639	&	4.520773733	\\\hline
15	&	0.029767	&	0.945226	&	0.949998	&	0	&	5024	&		0.043204	&	0.41474019	&	7.788255159	&	6.129831253	\\\hline
16	&	0.030184	&	0.932357	&	0.949996	&	0	&	18568	&		0.042452	&	0.395201719	&	8.37753314	&	6.719109234	\\\hline
17	&	0.030235	&	0.930807	&	0.949997	&	0	&	20200	&		0.042501	&	0.392843225	&	8.371020725	&	6.712596819	\\\hline
18	&	0.030251	&	0.930322	&	0.949998	&	0	&	20712	&		0.042284	&	0.392114741	&	8.500176807	&	6.841752902	\\\hline
19	&	0.028431	&	0	&	0.950198	&	800	&	0	&		0.045295	&	0.481217627	&	6.165721806	&	4.507297901	\\\hline
20	&	0.030509	&	0.922571	&	0.949999	&	0	&	28872	&		0.042212	&	0.380330252	&	8.64855253	&	6.990128624	\\\hline
21	&	0.030307	&	0.928628	&	0.95	&	0	&	22496	&		0.042723	&	0.389520645	&	8.275546956	&	6.617123051	\\\hline
22	&	0.030149	&	0.933423	&	0.949999	&	0	&	17448	&		0.042298	&	0.396829981	&	8.450038968	&	6.791615062	\\\hline
23	&	0.030778	&	0.914628	&	0.949998	&	0	&	37232	&		0.041647	&	0.368274421	&	9.085252889	&	7.426828984	\\\hline
24	&	0.030286	&	0.929263	&	0.949995	&	0	&	21824	&		0.042745	&	0.390484445	&	8.254117209	&	6.595693303	\\\hline
25	&	0.028312	&	0	&	0.950236	&	880	&	0	&		0.045654	&	0.487436318	&	5.936894638	&	4.278470733	\\\hline
26	&	0.029887	&	0.941485	&	0.949997	&	0	&	8960	&		0.042907	&	0.409065042	&	8.001486009	&	6.343062103	\\\hline
27	&	0.028648	&	0	&	0.950036	&	656	&	0	&		0.045069	&	0.46999291	&	6.350759686	&	4.692335781	\\\hline
28	&	0.028112	&	0	&	0.950226	&	1016	&	0	&		0.046306	&	0.498004738	&	5.524620815	&	3.866196909	\\\hline
29	&	0.030052	&	0.936392	&	0.949996	&	0	&	14320	&		0.04257	&	0.401332871	&	8.256950665	&	6.598526759	\\\hline
30	&	0.029944	&	0.939719	&	0.949995	&	0	&	10816	&		0.043148	&	0.406369206	&	7.889801849	&	6.231377943	\\\hline
31	&	0.028482	&	0	&	0.95007	&	768	&	0	&		0.045612	&	0.478548565	&	6.00545574	&	4.347031835	\\\hline
32	&	0.028341	&	0	&	0.950057	&	864	&	0	&		0.046096	&	0.485893688	&	5.699515506	&	4.041091601	\\\hline
33	&	0.028639	&	0	&	0.950331	&	656	&	0	&		0.045393	&	0.470441801	&	6.199128217	&	4.540704311	\\\hline
34	&	0.028151	&	0	&	0.950105	&	992	&	0	&		0.046136	&	0.495933886	&	5.617077248	&	3.958653342	\\\hline
35	&	0.028581	&	0	&	0.950301	&	696	&	0	&		0.045512	&	0.473424664	&	6.11261537	&	4.454191464	\\\hline
36	&	0.028332	&	0	&	0.950354	&	864	&	0	&		0.045976	&	0.486371377	&	5.783854353	&	4.125430447	\\\hline
37	&	0.028073	&	0	&	0.950348	&	1040	&	0	&		0.046308	&	0.500088447	&	5.520168261	&	3.861744355	\\\hline
38	&	0.029854	&	0.942511	&	0.949997	&	0	&	7880	&		0.043236	&	0.410610721	&	7.80502983	&	6.146605924	\\\hline
39	&	0.030517	&	0.922333	&	0.949996	&	0	&	29120	&		0.042385	&	0.379960638	&	8.5532151	&	6.894791194	\\\hline
40	&	0.028003	&	0	&	0.950325	&	1088	&	0	&		0.046221	&	0.503846816	&	5.539809861	&	3.881385956	\\\hline
41	&	0.028492	&	0	&	0.950129	&	760	&	0	&		0.0458	&	0.478021028	&	5.91422549	&	4.255801584	\\\hline
42	&	0.028991	&	0.970159	&	0.950375	&	416	&	0	&		0.044721	&	0.452593235	&	6.69144203	&	5.033018124	\\\hline
43	&	0.02827	&	0	&	0.950066	&	912	&	0	&		0.045812	&	0.489642361	&	5.825124611	&	4.166700705	\\\hline
44	&	0.030922	&	0.910432	&	0.949998	&	0	&	41648	&		0.041434	&	0.361903336	&	9.269858261	&	7.611434355	\\\hline
45	&	0.03083	&	0.913108	&	0.949998	&	0	&	38832	&		0.041848	&	0.365955157	&	8.991932879	&	7.333508973	\\\hline
46	&	0.029816	&	0.943695	&	0.949995	&	0	&	6632	&		0.043106	&	0.412416326	&	7.861992735	&	6.20356883	\\\hline
47	&	0.028779	&	0.977204	&	0.950375	&	560	&	0	&		0.045	&	0.463295864	&	6.463372889	&	4.804948983	\\\hline
48	&	0.030125	&	0.934156	&	0.949994	&	0	&	16672	&		0.042621	&	0.39793061	&	8.258044941	&	6.599621035	\\\hline

  \hline
\end{tabular}
}
\end{table}

\section*{Acknowledgements}

We would like to thank C. Ottaviani for the helpful discussions. This work was supported in part by the Key Program of National Natural Science Foundation of China under Grants 61531003, the National Natural Science Foundation under Grants 61427813, China Postdoctoral Science Foundation under Grant 2018M630116, and the Fund of State Key Laboratory of Information Photonics and Optical Communications.

%

\section*{Additional information}

\textbf{Competing financial interests:} The authors declare that they have no competing interests.

\end{document}